\pdfoutput=1

\documentclass[final,1p,times]{elsarticle} 
\usepackage{graphicx} 
\usepackage{amssymb} 
\usepackage{amsthm} 
\usepackage{lineno} 

\newcommand{\mean}[1]{\left\langle #1 \right\rangle}

\newcommand{\eps }{\varepsilon}
\def \vtep {\ensuremath{v_2\{\mathrm{EP}\} }}

\def \vtpp {\ensuremath{v_{2,\mathrm{PP}} }}

\def \vsep {\ensuremath{v\{\mathrm{subEP}\} }}
\def \vpp {\ensuremath{v_{\mathrm{PP}} }}

\def \npart {\ensuremath{N_{\mathrm{part}}}}

\def \sigtot {\ensuremath{\sigma_{\mathrm{tot}}}}

\def \epspart {\ensuremath{\eps_{\mathrm{part}}}}

\journal{Nuclear Physics A} 
\begin{document} 

\begin{frontmatter} 


\title{Effect of flow fluctuations and nonflow on elliptic flow methods\footnote{Condensed from arXiv:0904.2315 [nucl-ex]}}

\author{Jean-Yves Ollitrault, \underline{Arthur M. Poskanzer}, and Sergei A. Voloshin}

\address{CNRS, URA2306, Institut de physique th\'eorique de Saclay,
F-91191 Gif-sur-Yvette, France; Lawrence Berkeley National Laboratory, Berkeley,
California, 94720; Wayne State University, Detroit, Michigan, 48201}

\begin{abstract} 
We discuss how the different estimates of elliptic flow are influenced by flow fluctuations and nonflow effects. It is explained why the event-plane method yields estimates between the two-particle correlation methods and the multiparticle correlation methods. It is argued that nonflow effects and fluctuations cannot be disentangled without other assumptions. However, we provide equations where, with reasonable assumptions about fluctuations and nonflow, all measured values of elliptic flow converge to a unique mean $\vtpp$ elliptic flow in the participant plane. Thus, the 20\% spread in observed elliptic flow measurements from different analysis methods is no longer mysterious.
\end{abstract} 

\end{frontmatter} 


Elliptic flow has proved to be very valuable for understanding relativistic nuclear collisions~\cite{Voloshin:2008dg,Sorensen:2009}. However, different analysis methods give results which spread over a range of 20\%~\cite{Adams:2004bi}. A higher accuracy is now needed because when comparing to relativistic viscous hydrodynamic calculations, an uncertainty of 30\% in the elliptic flow parameter $v_2$ leads to an uncertainty of 100\% in the ratio of shear viscosity to entropy~\cite{Song:2008hj}.


For simplicity we will write $v\{\ \}$ instead of $v_n\{\ \}$ and $\cos(...)$ instead of $\cos[n(...)]$, where $n$ is the harmonic number of the anisotropic flow. The final equations are independent of $n$.

{\bf Flow Methods:}
The two-particle cumulant method $v\{2\}$ correlates each particle with every other particle, and is defined as
\begin{equation}
  v\{2\}\equiv\sqrt{\langle\cos(\phi_1-\phi_2)\rangle} \label{defv22} \ ,
\end{equation}
where $\mean{\ }$ indicates an average over all particles in all events.
The four-particle cumulant method $v\{4\}$ is defined as
\begin{equation}
  v\{4\}\equiv\left(2\langle\cos(\phi_1-\phi_2)\rangle^2-
  \langle\cos(\phi_1+\phi_2-\phi_3-\phi_4)\rangle\right)^{1/4}.
\label{defv24}
\end{equation}
The Lee-Yang Zeros method $v\{{\rm LYZ}\}$ is also a multiparticle correlation.
The event-plane estimate of anisotropic flow is defined as 
\begin{equation}
  v\{{\rm EP}\}\equiv \langle\cos(\phi-\Psi_R)\rangle / R \ ,
\label{defvep}
\end{equation}
where the particle of interest is always subtracted to avoid autocorrelations. $R$ is the event plane {\it resolution\/} correction which is determined from the correlation between the event plane vectors of two independent ``subevents'' $A$ and $B$. Methods of choosing the subevents are randomly, according to pseudorapidity or charge, or combinations of these. 

In the special case where the event plane comes from only one subevent the resolution correction is the subevent resolution
\begin{equation}
  R=\sqrt{\langle\cos(\Psi_A-\Psi_B)\rangle} \ . 
\label{subresolution}
\end{equation}
The corresponding estimate of anisotropic flow will be denoted by
\vsep, or, more particularly, $v\{{\rm etaSub}\}$ or $v\{{\rm ranSub}\}$, depending on how the events were divided. 

In the more general case when the event plane comes from the full event, one first estimates the resolution parameter $\chi_s$ of 
the subevents by solving numerically the equation
\begin{equation}
  {\cal R}(\chi_s)= \sqrt{\langle\cos(\Psi_A-\Psi_B)\rangle} \ ,
\label{defchis}
\end{equation}
where the function ${\cal R}$ is defined by~\cite{Poskanzer:1998yz,Ollitrault:1997di} 
\begin{equation}
  {\cal R}(\chi)=\frac{\sqrt{\pi}}{2}e^{-\chi^2/2}\chi\left(
  I_0\left(\frac{\chi^2}{2}\right)+I_1\left(\frac{\chi^2}{2}\right)
  \right),
\label{resolution}
\end{equation}
where $I_0$ and $I_1$ are modified Bessel functions. 
Generally, the resolution parameter is related to the flow through
$\chi_s=v\sqrt{N/2}$. One then estimates the resolution parameter $\chi$ of the full event as $\chi\equiv \chi_s\sqrt{2}$. The resolution correction for the full event $R$ is defined by $R\equiv {\cal R}(\chi)={\cal R}(\chi_s\sqrt{2})$.

{\bf Fluctuations:}
Elliptic flow is driven by the initial eccentricity of the overlap
almond~\cite{Ollitrault:1992bk}. This eccentricity fluctuates from one event to the other. It is fluctuations which make $\mean{v}$ in the participant plane larger than in the reaction plane. The magnitude of flow fluctuations is
characterized by $\sigma_v$, defined by 
\begin{equation}
\sigma_v^2\equiv\mean{v^2}-\mean{v}^2,
\label{defsigmav}
\end{equation} 
where $v$ is the flow in the participant plane $\vpp$ in the case of fluctuations in the participant plane. Flow methods involve various functions of $v$, which are also affected by fluctuations. 

We derive the effect of fluctuations on the various flow estimates, to order $\sigma_v^2$. Using the definitions of $v\{2\}$ and $v\{4\}$ from Eqs.~(\ref{defv22}) and (\ref{defv24}), 
\begin{equation}
v\{2\}^2= \mean{v^2}=\mean{v} ^2 +\sigma_v^2
\label{fluctv2}
\end{equation}
and
\begin{eqnarray}
v\{4\}^2&=& \left(2\mean{v^2}^2 -\mean{v^4}\right)^{1/2}
\approx \mean{v}^2-\sigma_v^2 \ .
\label{fluctv4}
\end{eqnarray}
Fluctuations increase $v\{2\}$ and decrease $v\{4\}$ compared to $\vpp$. 

{\bf Nonflow Effects:}
Now we discuss nonflow effects while neglecting fluctuations.
The two-particle azimuthal correlation gets contributions from flow
and from other ``nonflow'' effects 
\begin{equation}
  \mean{\cos(\phi_1-\phi_2)} \equiv \mean{v} ^2+\delta \ ,
\label{delDef}
\end{equation}
where $\delta$ is the nonflow part. One expects that $\delta$ varies with centrality like $1/N$, where $N$ is some measure of the multiplicity~\cite{Poskanzer:1998yz,Borghini:2000cm}. 

Using Eqs.~(\ref{defv22}) and (\ref{delDef}), one obtains, to leading
order in $\delta$ 
\begin{equation}
v\{2\}^2=\mean{v}^2+\delta \ .
\label{nonflowv2}
\end{equation}
On the other hand, $v\{4\}$ is insensitive to nonflow effects, and thus
\begin{equation}
v\{4\}=\mean{v} \ .
\label{nonflowv4}
\end{equation}

{\bf Equations:}
We assume that to leading order in $\sigma_v^2$ and $\delta$, the contributions of nonflow and fluctuations are additive. Eqs.~(\ref{fluctv2}) and (\ref{nonflowv2}) yield
\begin{equation}
Êv\{2\}^2=
Ê\langle v\rangle^2 +\delta+\sigma_v^2 \ .
\label{sumv2}
\end{equation}
Similarly, Eqs.~(\ref{fluctv4}) and (\ref{nonflowv4}) yield 
\begin{equation}
Êv\{4\}^2=
Ê\langle v\rangle^2 -\sigma_v^2 \ .
\label{sumv4}
\end{equation}
Although this equation was derived for $v\{4\}$ it should apply to all multiparticle values. As for the event-plane methods, one can show that
\begin{equation}
Êv\{{\rm EP}\}^2=\langle
Êv\rangle^2+
Ê\left(1-\frac{(I_0-I_1)}{(I_0+I_1)}\left(\chi^2-\chi_s^2+\frac{2 i_1^2}{
Ê(i_0^2-i_1^2)}\right)\right)\delta
Ê+\left(1-\frac{2(I_0-I_1)}{I_0+I_1}\left(\chi^2-\chi_s^2+\frac{2
Êi_1^2}{i_0^2-i_1^2}\right)\right)\sigma_v^2
\label{sumvep}
\end{equation}
\begin{equation}
Êv\{{\rm subEP}\}^2= \langle v\rangle^2 +
Ê\left(1-\frac{2 i_1^2}{
Ê(i_0+i_1)^2}\right)\delta
Ê+\left(1-\frac{4\,i_1^2}{(i_0+i_1)^2} \right)
Ê\sigma_v^2 \ , 
\label{sumvsub}
\end{equation}
where $i_{0,1}$ is a shorthand notation for $I_{0,1}(\chi_s^2/2)$ and $I_{0,1}$ for $I_{0,1}(\chi^2/2)$. The differences between the various $v_2\{\ \}$ estimates always scale like $\delta+2\sigma_v^2$. This shows explicitly that fluctuations and nonflow effects cannot be disentangled with only these measurements. Thus we have defined $\sigma_{\rm tot}^2 \equiv \delta + 2 \sigma_{v}^2$.

{\bf Application to Data:}
So far the equations have used generic fluctuation and nonflow parameters. To apply the analytic equations to extract $\mean{v}$ in the participant plane from experimental data, we have assumed that the fluctuations in $v$ have the same fractional width as the fluctuations of the participant eccentricity
\begin{equation}
  \sigma_{v} = (\sigma_{\varepsilon} / \langle \varepsilon \rangle) \ \langle v \rangle \ .
\label{sigma_v}
\end{equation}
A nucleon Monte-Carlo Glauber calculation was used to calculate the fractional standard deviation of \epspart~\cite{Hiroshi}. For the nonflow contribution we have taken the value from proton-proton collisions and scaled it down by the number of participants. The value of $\delta_{pp}$ was obtained by integrating the minimum bias $p+p$ curves of Ref.~\cite{Adams:2004wz}, Fig.~1, and it was found that $\delta_{pp} = 0.0145$~\cite{Tang}. Thus for nonflow as a function of centrality we assume 
\begin{equation}
  \delta = \delta_{pp} \ 2 / \npart \ ,
\label{delta}
\end{equation}
knowing that in a $p+p$ collision there are two participants. 

\begin{figure*}[hbt]
\begin{minipage}[t]{0.43\textwidth}
 \includegraphics[width=0.98\textwidth]{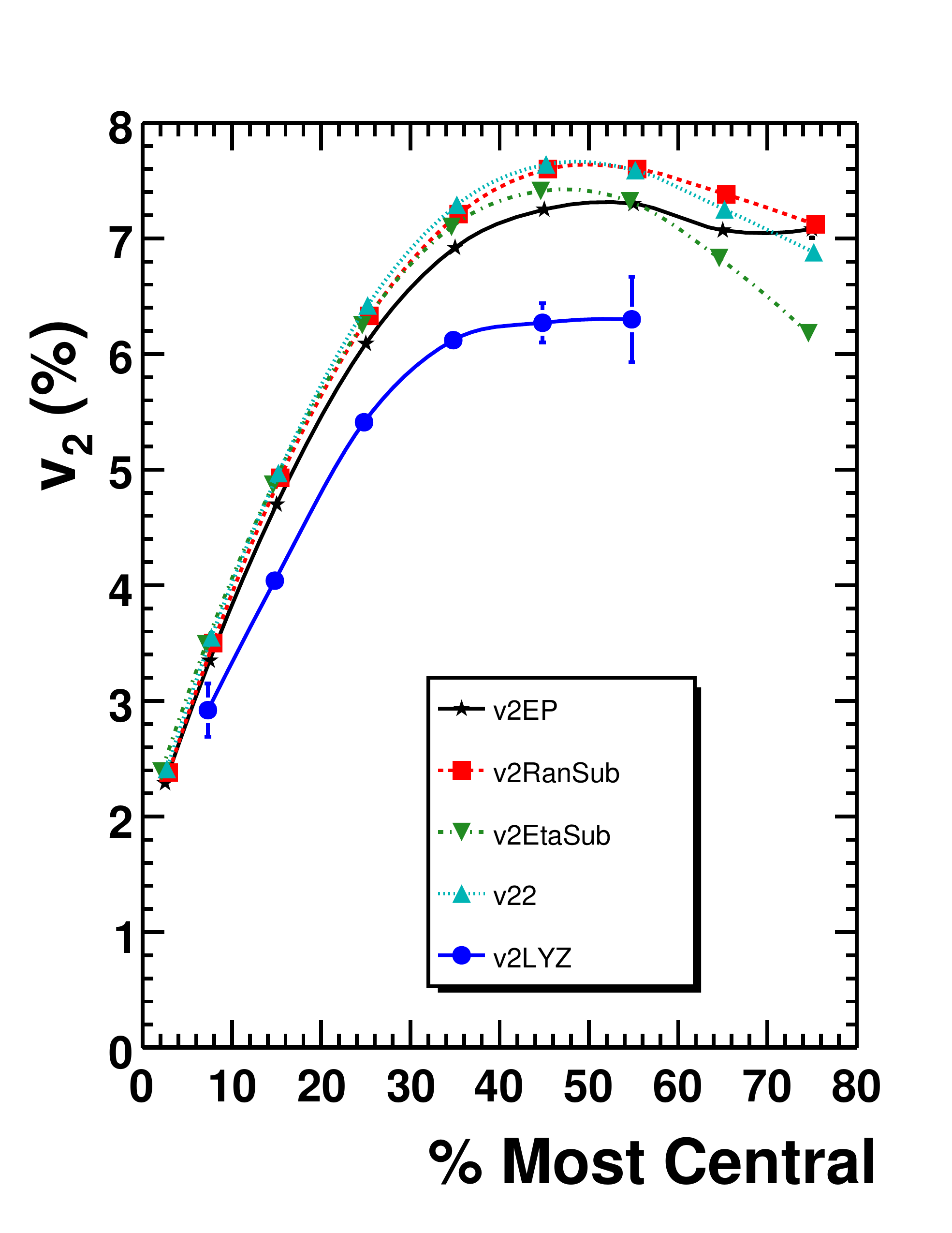}
 \caption{(Color online) The values of $v_2$ from various analysis methods vs centrality. Both the upper lines~\cite{Adams:2004bi} and the lower line~\cite{Abelev:2008ed} are STAR data. }
 \label{fig:pub_v}
\end{minipage}
\hspace{\fill}
\begin{minipage}[t]{0.43\textwidth}
\includegraphics[width=0.98\textwidth]{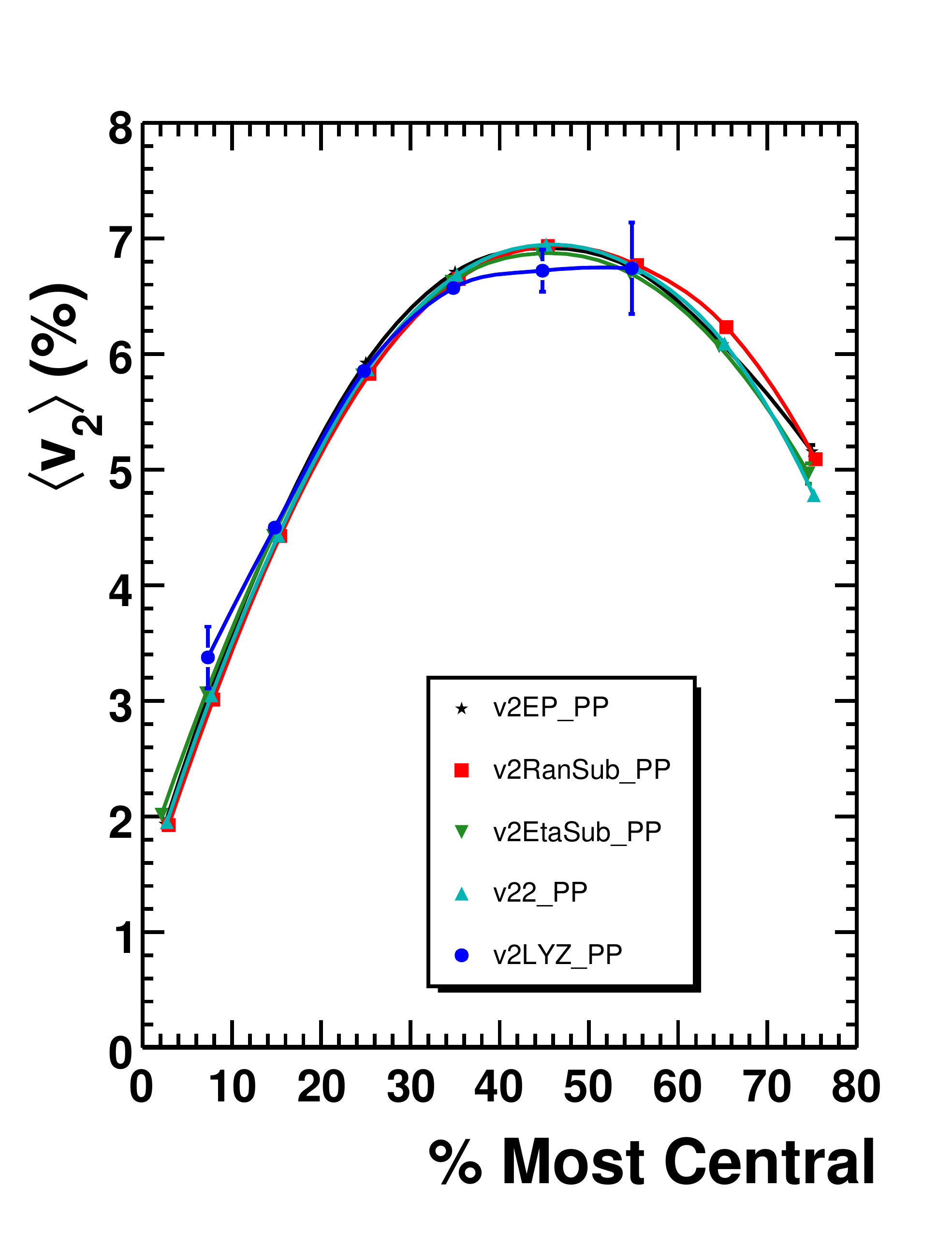}
 \caption{(Color online) The data from Fig.~\ref{fig:pub_v} corrected to $\mean{v_2}$ in the participant plane.}
  \label{fig:mean_v}
\end{minipage}
\end{figure*}
The published STAR data~\cite{Adams:2004bi,Abelev:2008ed} for the various methods are shown in Fig.~\ref{fig:pub_v}. The upper lines are from ``two-particle" correlation methods, and the lower line is from a multiparticle correlation method. The lower line values for $v_2\{{\rm LYZ}\}$ are thought to be in the reaction plane, if the fluctuations are Gaussian~\cite{Voloshin:2007pc}. The line for $v_2\{{\rm etaSub}\}$ is somewhat low for peripheral collisions because the gap in pseudorapidity reduces short-range nonflow correlations. Particularly puzzling is why the $\vtep$ line is lower than the other two-particle methods.

Correcting to $\mean{v}$ in the participant plane was done by using Eq.~(\ref{sumv2}) for $v_2\{2\}$, Eq.~(\ref{sumv4}) for $v_2\{{\rm LYZ}\}$, Eq.~(\ref{sumvep}) for \vtep, and Eq.~(\ref{sumvsub}) for $v_2\{{\rm ranSub}\}$ and $v_2\{{\rm etaSub}\}$. The results are shown in Fig.~\ref{fig:mean_v}. Since $v_2\{{\rm etaSub}\}$ is less affected by nonflow, the value of $\delta_{pp}$ used for it was multiplied by 0.5. In Fig.~\ref{fig:mean_v} the convergence of the two-particle, full event plane, and multiparticle results to one locus in the participant plane is remarkable. Even the shape of the $v_2\{{\rm etaSub}\}$ curve has changed to match the others with only one additional parameter. Previously we took the spread in the values in Fig.~\ref{fig:pub_v} as an estimate of the systematic uncertainty.

{\bf Summary:}
We have shown how the various experimental measures of elliptic flow are affected by fluctuations and nonflow, and we derived analytic equations which are leading order in $\sigma_v^2$ and $\delta$. We have transformed published data to the participant plane using reasonable assumptions for fluctuations and nonflow. The convergence of the various experimental measurements is remarkable. The convergence of the methods essentially fixes the value of $\sigtot$ from experimental data, but the separation into fluctuation and nonflow parts is not unique. To avoid both, better results for multiparticle correlations are needed. 



\begin{thebibliography}{00} 
   
\bibitem{Voloshin:2008dg}
  S.~A.~Voloshin, A.~M.~Poskanzer and R.~Snellings,
  arXiv:0809.2949 [nucl-ex].

\bibitem{Sorensen:2009}
  P.~Sorensen,
  arXiv:0905.0174 [nucl-ex].
  
\bibitem{Adams:2004bi}
 J.~Adams {\it et al.}  [STAR Collaboration],
 Phys.\ Rev.\  C {\bf 72}, 014904 (2005)
 [arXiv:nucl-ex/0409033].

\bibitem{Song:2008hj}
  H.~Song and U.~W.~Heinz,
  arXiv:0812.4274 [nucl-th].
  
\bibitem{Poskanzer:1998yz}
 A.~M.~Poskanzer and S.~A.~Voloshin,
 Phys.\ Rev.\  C {\bf 58}, 1671 (1998)
 [arXiv:nucl-ex/9805001].

\bibitem{Ollitrault:1997di}
  J.~Y.~Ollitrault,
  arXiv:nucl-ex/9711003; Nucl.\ Phys.\  A {\bf 638}, 195 (1998).
  [arXiv:nucl-ex/9802005].
  
\bibitem{Ollitrault:1992bk}
 J.~Y.~Ollitrault,
 Phys.\ Rev.\  D {\bf 46}, 229 (1992).

\bibitem{Borghini:2000cm}
 N.~Borghini, P.~M.~Dinh and J.~Y.~Ollitrault,
 Phys.\ Rev.\  C {\bf 62}, 034902 (2000)
 [arXiv:nucl-th/0004026].

\bibitem{Hiroshi}
 Hiroshi Masui, private communication, 2008.
 
\bibitem{Adams:2004wz}
 J.~Adams {\it et al.}  [STAR Collaboration],
 Phys.\ Rev.\ Lett.\  {\bf 93}, 252301 (2004)
 [arXiv:nucl-ex/0407007].

\bibitem{Tang}
 Aihong Tang, private communication, 2008.

\bibitem{Abelev:2008ed}
  B.~I.~Abelev {\it et al.}  [STAR Collaboration],
  Phys.\ Rev.\  C {\bf 77}, 054901 (2008)
  [arXiv:0801.3466 [nucl-ex]].

\bibitem{Voloshin:2007pc}
 S.~A.~Voloshin, A.~M.~Poskanzer, A.~Tang and G.~Wang,
 Phys.\ Lett.\  B {\bf 659}, 537 (2008)
 [arXiv:0708.0800 [nucl-th]].

\end{thebibliography}
\end{document}